\documentclass[11pt,sort&compress]{elsarticle}
\usepackage[paper=letterpaper,top=24mm, bottom=26mm, left=26mm, right=26mm]{geometry}
\makeatletter
\def\ps@pprintTitle{%
 \let\@oddhead\@empty
 \let\@evenhead\@empty
 \def\@oddfoot{\centerline{\thepage}}%
 \let\@evenfoot\@oddfoot}
\makeatother

\usepackage[hyperref]{xcolor}
\definecolor{darkgreen}{rgb}{0.01, 0.75, 0.24}
\usepackage[colorlinks=true,
            linkcolor=darkgreen,
            urlcolor=darkgreen,
            citecolor=darkgreen,linkcolor=darkgreen,hyperfootnotes=true]{hyperref}
\usepackage{amsmath}
\usepackage{amsfonts}
\usepackage{slashed}
\usepackage{accents}
\usepackage{amssymb}
\usepackage{mathrsfs}
\usepackage{mathtools}
\usepackage[LGR,T1]{fontenc}
\usepackage[latin1]{inputenc}
\let\oldbibliography\thebibliography
\renewcommand{\thebibliography}[1]{%
  \oldbibliography{#1}%
  \setlength{\itemsep}{1.4pt}%
}

\usepackage[cal=boondox,calscaled=1]{mathalfa}
\DeclareMathAlphabet{\bbvar}{U}{BOONDOX-ds}{m}{n}
\DeclareMathAlphabet{\bbgreek}{U}{bbold}{m}{n}
\usepackage{psfrag}
\usepackage{pstool}
\usepackage{caption}
\usepackage{tabularx}
\usepackage{multirow}
\usepackage{pbox}
\usepackage{graphicx}
\newcommand{\hook}{\text{\large{$\lrcorner$}}}

\newcommand{\darkg}[1]{{\color{darkgreen}#1}}

\newcommand{\qq}[1]{``#1''} 

\newcommand{\di}{\mathrm{d}}
\usepackage{tensor}
\newcommand{\ou}[3]{\tensor{#1}{^{#2}_{#3}}}

\newcommand{\R}{\mathbb{R}}

\newcommand{\eref}[1]{(\ref{#1})}

\newcommand{\bbwedge}{\reflectbox{\rotatebox[origin=c]{180}{\fontsize{10pt}{10pt}$\hspace{0.7pt}\bbvar{V}\hspace{0.7pt}$}}}
\newcommand{\bbhook}{\text{\large{$\lrcorner$}\hspace{-0.40em}\large{$\lrcorner$}}}

\usepackage{tikz-cd}

\newcommand\vpm{\mathbin{\vcenter{\hbox{
  \oalign{\hfil$\scriptstyle+$\hfil\cr
          \noalign{\kern-.3ex}
          $\scriptscriptstyle({-})$\cr}}}}}
\DeclareMathAlphabet{\sfit}{OT1}{fos}{sb}{it}
\DeclareMathAlphabet{\mathsf}{OT1}{fos}{sb}{n}

\definecolor{darkgreen}{rgb}{0.01, 0.75, 0.24}
\definecolor{darkblue}{rgb}{0.01, 0.24, 0.75}

\usepackage[multiple, flushmargin]{footmisc}

\let\originalleft\left
\let\originalright\right
\renewcommand{\left}{\mathopen{}\mathclose\bgroup\originalleft}
\renewcommand{\right}{\aftergroup\egroup\originalright}

\newcommand{\llbrack}{\{\hspace{-0.23em}|}
\newcommand{\rrbrack}{|\hspace{-0.23em}\}}

\newcommand{\dbarvar}{{\mathrm{d}\mkern-7.5mu\lower.18ex\hbox{$\textasciitilde$}\mkern-1.5mu}}

\renewcommand{\emph}[1]{{\it #1}}

\begin{document}

\begin{abstract}
\noindent In general relativity, it is difficult to localise observables such as energy, angular momentum, or centre of mass in a bounded region. The difficulty is that there is dissipation. A self-gravitating system, confined by its own gravity to a bounded region, radiates some of the charges away into the environment. At a formal level, dissipation implies that some diffeomorphisms are not Hamiltonian. In fact, there is no Hamiltonian on phase space that would move the region \emph{relative} to the fields. Recently, an extension of the covariant phase space has been introduced to resolve the issue. On the extended phase space, the Komar charges are Hamiltonian. They are generators of \emph{dressed diffeomorphisms}. While the construction is sound, the physical significance is unclear. We provide a critical review before developing a geometric approach that takes into account dissipation in a novel way. Our approach is based on metriplectic geometry, a framework used in the description of dissipative systems. Instead of the Poisson bracket, we introduce a \emph{Leibniz bracket}---a sum of a skew-symmetric and a symmetric bracket. The symmetric term accounts for the loss of charge due to radiation. On the metriplectic space, the charges are Hamiltonian, yet they are not conserved under their own flow.

\end{abstract}%
\title{Metriplectic geometry for gravitational subsystems}
\author{Viktoria Kabel${}^{1,2}$ and Wolfgang Wieland${}^{1,2}$}
\address{${}^{1}$Institute for Quantum Optics and Quantum Information (IQOQI)\\Austrian Academy of Sciences\\Boltzmanngasse 3, 1090 Vienna, Austria
}
\address{${}^{2}$Vienna Center for Quantum Science and Technology (VCQ)\\Faculty of Physics, University of Vienna\\ Boltzmanngasse 5, 1090 Vienna, Austria\\{\vspace{0.5em}\normalfont 30 May 2022}}
\maketitle
\vspace{-1.2em}
\hypersetup{
  linkcolor=black,
  urlcolor=black,
  citecolor=black
}
{\tableofcontents}
\hypersetup{
  linkcolor=black,
  urlcolor=darkgreen,
  citecolor=darkgreen,
}
\begin{center}{\noindent\rule{\linewidth}{0.4pt}}\end{center}\newpage
\section{The problem considered}{
\noindent 
Consider a region  of space with fixed initial data.  What is the total energy contained in the region? General relativity gives no definite answer to this question. There is no unique \emph{quasi-local} \cite{Szabados:2004vb,BrownLauYork,BrownYork} notion of energy in general relativity. This is due to two features of the theory: first of all, there are no preferred coordinates. If there are no preferred coordinates, there is 
 no preferred notion of time. Time is dual to energy. If there is no preferred clock \cite{Rovelli:2009ee}, there is also no preferred notion of energy. The second issue is dissipation. If we insist to restrict ourselves to local observables in a finite region of space, we have to specify what happens at the boundary. Since gravity can not be shielded, there is always dissipation. A local gravitational system will always be open. Gravitational radiation carries away gravitational charge, including mass, energy, angular momentum, centre of mass, and additional soft modes related to gravitational memory \cite{PhysRevLett.67.1486,memory_frauendiener,Hawking:2016msc}, which makes it difficult to characterise gravitational charges on the full non-perturbative phase space of the theory.\medskip

That there is no preferred notion of energy or momentum does not mean, of course, that it would be impossible to speak about such important physical concepts in general relativity. One possibility to do so is to introduce material frames of reference $\{X^\mu\}$, which depend themselves---in  a highly non-linear but covariant way\footnote{The condition is
$
\forall\phi\in\mathrm{Diff}(M:M), p\in M : X^\mu[g_{ab},\psi^I]\big(\phi(p)\big) = X^\mu[\phi^\ast g_{ab},\phi^\ast\psi^I](p).
$}---on the metric $g_{ab}$ and the matter fields $\psi^I$.
 The resulting \emph{dressed} observables, a version of Rovelli's \emph{relational observables} \cite{Rovelli_1991,PhysRevD.42.2638,Rovelli2001,Thiemann:2004wk,Dittrich:2005kc,Tambornino:2011vg}, evaluate the kinematical observables at those events in spacetime, where the physical frames of reference take a certain value. Such a dressing  turns a gauge dependent kinematical observable, such as the metric, into a  gauge invariant (Dirac) observable \cite{Donnelly:2016rvo,Donnelly:2018nbv}. An example for such an observable is the dressed metric itself, 
\begin{equation}
g^{\mu\nu}\big[g_{ab},\psi^I\big](x_o) = \int_M \di X^0\wedge\dots\wedge\di X^3\,\delta^{(4)}\big(X^\mu-x^\mu_o\big)\,g^{ab}\partial_aX^\mu\partial_bX^\nu,\label{relational1}
\end{equation}
where the reference frame $X^\mu$ itself depends functionally on metric and matter fields, i.e.\ $X^\mu\equiv X^\mu[g_{ab},\psi^I]$. 
Given a material reference frame  $X^\mu[g_{ab},\psi^I]$, we have a natural class of state-dependent vector fields $\xi^a=\xi^\mu\big(X[g_{ab},\psi^I]\big)\big[\tfrac{\partial}{\partial X^\mu}\big]^a$. 
On shell, the corresponding Hamiltonian, {so it exists}, defines a surface charge $Q_\xi$, which is  conjugate to the reference frame, i.e.\ 
\begin{equation}\left.
\begin{split}
\big\{Q_{\xi},g_{ab}\big\}&=\mathcal{L}_{\xi}g_{ab},\\
\big\{Q_{\xi},\psi^I\big\}&=\mathcal{L}_{\xi}\psi^I
\end{split}
\right\}
\Longrightarrow \big\{Q_{\xi},X^\mu\big\}=\xi^\mu(X),\label{relational2}
\end{equation}
where $\mathcal{L}_\xi$ is the Lie derivative. That the Hamiltonian is a surface charge is a consequence of Noether's theorem and the diffeomorphism invariance of the action. While this approach is intuitive, it is unpractical. It is unpractical, because the construction depends for any realistic choice of coordinates $\{X^\mu[g_{ab},\psi^I]\}$ on the metric and matter fields in a complicated and highly non-local way \cite{Torre:1993fq}. A further difficulty is that there are no such coordinates defined globally on the entire state space.\medskip
 
 A more practical approach is to take advantage of asymptotic boundary conditions. In an asympotically flat spacetime,  the asymptotic boundary conditions select a specific class of asymptotic (BMS) coordinates \cite{PhysRev.128.2851,Sachs103}. Any two members of this class can be mapped into each other via an asymptotic symmetry, generated by an asymptotic BMS vector field $\xi^a_{\mathrm{BMS}}$. One may then  expect that 
there is a corresponding charge $Q_{\xi_{\mathrm{BMS}}}$ that would generate the asymptotic symmetry as a motion on phase space. This, however, immediately leads to the second problem mentioned above: dissipation. The system is open, because radiation escapes to null infinity, and the charges cannot be conserved under their own flow.
Hence, the BMS charges cannot be Hamiltonian, i.e.\ there is no charge on a two-dimensional cross section of future (past) null infinity that would generate an asymptotic symmetry, see also \cite{AshtekarNullInfinity} for a more detailed explanation of this issue on the radiative phase space. \medskip
  
The same issues appear also at finite distance \cite{Donnelly:2016auv,Donnelly:2017jcd,Harlow:2020aa,Wieland:2017zkf,Wieland:2021vef,Wieland:2020gno}. A candidate for a quasi-local notion of gravitational energy in a finite region $\Sigma$, often mentioned in the literature, is the Komar charge \cite{PhysRev.113.934}. On the usual covariant phase space \cite{Lee:1990nz,Ashtekar:1990gc,Wald:1999wa}, it is not at all obvious what the Hamiltonian vector field of the Komar charges should be. The naive expectation that would identify the Komar charge with the generator of a diffeomorphism is incorrect. It is incorrect, because there is dissipation. The charges cannot be conserved under their own flow. This, at least, is the usual story. \medskip

}

Recently, a different viewpoint appeared on the issue of dissipation and Hamiltonian charges. The basic idea put forward by Ciambelli, Leigh, Pai \cite{Ciambelli:2021vnn,PhysRevLett.128.171302}, Freidel and collaborators \cite{Freidel:2021dxw,Freidel:2021cjp} and Speranza and Chandrasekaran  \cite{Speranza:2017gxd,Chandrasekaran:2020wwn} is to add boundary modes and extend the covariant phase space in such a way that the Komar charges become Hamiltonian. The resulting modified Poisson bracket on phase space returns the Barnich--Troessaert bracket \cite{Barnich:2011mi,Barnich:2010eb,Wieland:2021eth} between the charges. These ideas resulted from a wider research programme concerned with 
gravitational subsystems, quasi-local observables,  physical reference frames, deparametrisation, and the meaning of gauge \cite{Rovelli:2013fga,Gomes:2019xhu,Donnelly:2016auv,Donnelly:2017jcd,Speranza:2017gxd,Harlow:2020aa,Freidel:2020xyx,Wieland:2017zkf,Wieland:2021vef,Wieland:2017cmf,Wieland:2020gno,Freidel:2020xyx,Chandrasekaran:2020wwn,Gomes:2016mwl,Gomes:2019xto,Dittrich:2018xuk,Margalef-Bentabol:2020teu,Carrozza:2022xut}.\medskip

In the following, we shall give a concise and critical summary (\hyperref[sec2]{\darkg{Section 2}}) of the construction \cite{Ciambelli:2021vnn,PhysRevLett.128.171302,Freidel:2021dxw} before developing a more geometric \emph{metriplectic approach} \cite{Morrison86,BKMR96,Fish2005} in \hyperref[sec3]{\darkg{Section 3}}. In the metriplectic approach, the usual Poisson bracket is replaced by a Leibniz bracket on covariant phase space. This new bracket consists of a symmetric and a skew-symmetric part. The skew-symmetric part defines a Poisson bracket on the extended phase space. The symmetric part captures dissipation. Some of the charge aspect is carried away under the Hamiltonian flow into the environment. For a gravitational system, restricted to a bounded region of space, the Komar charges are canonical with respect to the Leibniz bracket. The charges  generate diffeomorphisms of the region \emph{relative} to the fields inside. They are Hamiltonian, but are not conserved under their own Hamiltonian flow, thus accounting for dissipation in gravitational subsystems.

\section{Dressing and covariant phase space}\label{sec2}

\subsection{Extended symplectic structure}
\noindent The starting point of the original dressed phase space approach due to \cite{PhysRevLett.128.171302} and \cite{Freidel:2021dxw} is the usual state space of general relativity consisting of solutions to the Einstein equations  $R_{ab}[g]-\tfrac{1}{2}g_{ab}R[g]=8\pi G\,T_{ab}[g_{ab},\psi^I]$ for a metric $g_{ab}$, and some matter fields $\psi^I$ on an abstract and differentiable manifold $M$. The state space $\mathcal{F}\ni(g_{ab},\psi^I)$ is then extended by including a gravitational dressing for the diffeomorphism group. A point on the extended state space $\mathcal{F}_{\mathit{ext}}\ni(g_{ab},\psi^I,\phi)$ is thus characterised by a solution $(g_{ab},\psi^I)$ to Einstein's equations on $M$ and a diffeomorphism $\phi:M\rightarrow M$, which is purely kinematical.\footnote{This is to say that there are no field equations or gauge conditions that would constrain $\phi$.} The diffeomorphism, which has now been added to state space, allows us to introduce dressed solutions to the Einstein's equations, i.e.\ $(\phi^\ast g_{ab},\phi^\ast\psi^I)$, where $\phi^\ast$ denotes the pull-back. 

At first, the construction seems to merely add further redundancy and to run against our basic physical intuition about background invariance.
In a generally covariant theory, a diffeomorphism should have no physical meaning whatsoever and $(g_{ab},\psi^I)$ ought to represent the same physical state as $(\phi^\ast g_{ab},\phi^\ast\psi^I)$. But this intuition is  slightly misleading. It is misleading for two reasons. The first reason is that boundaries break gauge symmetries, turning otherwise redundant gauge directions into physical boundary modes. If $\phi:M\rightarrow M$ is a large diffeomorphism such that $\phi|_{\partial M}\neq\mathrm{id}$, the two states $(\phi^\ast g_{ab},\phi^\ast\psi^I)$ and $(g_{ab},\psi^I)$ are no longer gauge equivalent (in the phase space sense of the word). The second reason is that the extended symplectic potential proposed in \cite{PhysRevLett.128.171302} and \cite{Freidel:2021dxw} has a highly-non trivial dependence on $\phi$. The gravitational dressing $\phi^\ast$ enters the extended pre-symplectic current $\vartheta_{\mathit{ext}}$ through two independent terms
\begin{equation}
\vartheta_{\mathit{ext}} = \phi^\ast\vartheta + \phi^\ast(\bbvar{Y}\hook L),\label{thetaext}
\end{equation}
where $L$ is the Lagrangian, which, in turn, defines the pre-symplectic current\footnote{In the following, all equations are taken on-shell, i.e.\ provided the field equations are satisfied.}
\begin{equation}
\forall\delta\in T\mathcal{F}:\delta[L]=\di\big[\vartheta(\delta)\big].
\end{equation}
In addition, the extended pre-symplectic current depends on $\bbvar{Y}^a$, which is a $TM$-valued one-form on the extended state space $\mathcal{F}_{\mathit{ext}}$, i.e.\ a section of the tensor bundle $TM\otimes T^\ast\mathcal{F}_{\mathit{ext}}$, and behaves like a Maurer--Cartan form $(\bbvar{d}\phi)(\phi^{-1})$ for diffeomorphisms. 

The one-form $\bbvar{Y}^a$ on field space can be introduced as follows. Consider an ordinary state-independent differentiable function on spacetime, say $f:M\rightarrow \R$. Since (spacetime) vector fields are derivations acting on scalars, the expression $\bbvar{Y}_p[f]\equiv\bbvar{Y}^a\partial_af\big|_p$  must be read as a one-form on field space, i.e.\ for all $p\in M:\bbvar{Y}_p[f]\in T^\ast\mathcal{F}_{\mathit{ext}}$. This one-form, which will depend linearly on $\di f\in T^\ast M$, is itself defined by
\begin{equation}
\bbvar{Y}_p[f] := \big(\bbvar{d}(f\circ\phi)\big)(\phi^{-1}(p))\equiv\bbvar{Y}^a_p (\partial_a f)_p,\label{Ydef1}
\end{equation}
where the symbol \qq{$\bbvar{d}$} denotes the exterior derivative on field space $\mathcal{F}_{\mathit{ext}}\ni(g_{ab},\psi^I,\phi)$. An explicit coordinate expression of $\bbvar{Y}^a$ with respect to some fixed and fiducial coordinate chart $\{x^\mu\}$ in a neighbourhood of $p\in M$ is thus given by
\begin{equation}
\bbvar{Y}^a\big|_p = \bbvar{d}[x^\mu\circ \phi]\big|_{\phi^{-1}(p)} \Big[\frac{\partial}{\partial x^\mu}\Big]^a_p.\label{Ydef2}
\end{equation}
In the following, let us study this one-form a little more carefully. If we commute the field space derivative with the dressing, we obtain
\begin{equation}
\bbvar{d}\phi^\ast = \phi^\ast\bbvar{d} +\phi^\ast\mathcal{L}_{\bbvar{Y}}.
\end{equation}
This equation is obviously true for scalars. The generalisation to arbitrary $p$-form fields is immediate and is the consequence of two basic observations: the exterior derivative (on spacetime) commutes with the pull-back, i.e.\ $\phi^\ast\di=\di\phi^\ast$, and the exterior derivative on field space commutes with the exterior derivative on spacetime, i.e.\ $[\di,\bbvar{d}]=0$. 

The one-form $\bbvar{Y}^a$ behaves like a Maurer--Cartan form for the diffeomorphism group  \cite{PhysRevLett.128.171302,Gomes:2016mwl}. If, in fact, $\delta_1$ and $\delta_2$ are two tangent vectors on field space, equation  \eref{Ydef2} immediately implies
\begin{align}
\big(\bbvar{d}\bbvar{Y}^a\big)(\delta_1,\delta_2)  = - [\bbvar{Y}(\delta_1),\bbvar{Y}(\delta_2)]^a.
\end{align}
In other words,
\begin{equation}
\bbvar{d}\bbvar{Y}^a =- \bbvar{Y}^b\bbwedge\nabla_b\bbvar{Y}^a,
\end{equation}
where $\nabla_a$ is the metric compatible torsionless derivative with respect to $g_{ab}$, i.e.\ $\nabla_ag_{bc}=0$ and $\nabla_{[a}\nabla_{b]}f=0,\,\forall f:M\rightarrow\R$.
\medskip

For a background invariant theory, it is now always possible to trivially absorb the dressing field $\phi\in\mathrm{Diff}(M:M)$ back into a redefinition of metric and matter fields. We shall find  this redefinition useful, because it will clarify the physical significance of the construction. If, in fact, the theory is background invariant, the symplectic current transforms covariantly under diffeomorphisms. In other words,
\begin{equation}
\forall p\in M:\big(\phi^\ast\vartheta[g_{ab},\psi^I;\bbvar{d}g_{ab},\bbvar{d}\psi^I]\big)(p) = \vartheta[\phi^\ast g_{ab},\phi^\ast\psi^I;\phi^\ast\bbvar{d}g_{ab},\phi^\ast\bbvar{d}\psi^I](p).\label{covtheta}
\end{equation}
On the other hand, we also have
\begin{align}
\phi^\ast\bbvar{d}g_{ab} & = \bbvar{d}(\phi^\ast g_{ab}) - \phi^\ast \mathcal{L}_{\bbvar{Y}} g_{ab},\label{commut1}\\
\phi^\ast\bbvar{d}\psi^I & = \bbvar{d}(\phi^\ast \psi^I) - \phi^\ast \mathcal{L}_{\bbvar{Y}} \psi^I,\label{commut2}
\end{align}
where $\mathcal{L}_{\bbvar{Y}}$ denotes the Lie derivative with respect to $\bbvar{Y}^a\in T^\ast\mathcal{F}_{\mathit{ext}}\otimes TM$, i.e.\
\begin{equation}
\mathcal{L}_{\bbvar{Y}}[g_{ab}]=2\nabla_{(a}{\bbvar{Y}}_{b)},\quad\mathcal{L}_{\bbvar{Y}}\psi^I=\frac{\di}{\di\varepsilon}\Big|_{\varepsilon=0}\exp(\varepsilon\,{\bbvar{Y}})^\ast\psi^I.
\end{equation}

A trivial field redefinition allows us to remove the dressing fields and absorb them back into the definition of metric and matter fields
\begin{align}
\phi^\ast g_{ab}&\longrightarrow g_{ab},\qquad
\phi^\ast \psi^I\longrightarrow\psi^I,\label{fieldredef}\\
\phi^{-1}_\ast\bbvar{Y}^a&= \bbvar{d}(x^\mu\circ\phi)\Big[\phi^{-1}_{\ast}\frac{\partial}{\partial x^\mu}\Big]^a=\bbvar{d}(x^\mu\circ\phi)\Big[\frac{\partial}{\partial (x^\mu\circ\phi)}\Big]^a=:\bbvar{X}^a.\label{Xdef}
\end{align}
So far, we kept the fiducial coordinate system fixed $\{x^\mu\}$  and treated the diffeomorphisms $\phi:M\rightarrow M$ as a new dynamical element of the thus extended state space $\mathcal{F}_{\mathit{ext}}$. Equation \eref{Xdef} tells us that we could also adopt a different viewpoint. Instead of adding the diffeomorphism to the state space, we could equally well reabsorb the dressing $\phi^\ast$ into a redefinition of the coordinates, i.e.\ $\phi^\ast x^\mu=x^\mu\circ\phi\longrightarrow x^\mu$.  Adopting this viewpoint amounts to adding the four coordinate scalars $x^\mu:M\rightarrow \R^4$ to the state space.

The addition of coordinate functions to phase space seems to run against the very idea of background invariance. To restore formal coordinate invariance, it is useful to introduce the Maurer--Cartan form
\begin{equation}
\bbvar{X}^a = \bbvar{d}[x^\mu]\,\partial^a_\mu,
\end{equation}
which sends the coordinate variations $\delta x^\mu=\delta\bbhook\bbvar{d} x^\mu$ back into tangent space $TM$. Once again, this one-form on field space behaves like a ghost field for the diffeomorphism group, i.e.\
\begin{align}\nonumber
\bbvar{d}\bbvar{X}^a & = - \bbvar{d}x^\mu \bbwedge \bbvar{d}\big[\partial^a_\mu\big] = 
 - \bbvar{d}x^\mu \bbwedge \partial_\mu^b\di x_b^\nu\bbvar{d}\big[\partial^a_\nu\big]= + \bbvar{d}x^\mu \bbwedge \partial_\mu^b\bbvar{d}\big[\di x_b^\nu\big]\partial^a_\nu=\\
 &= + \bbvar{d}x^\mu \bbwedge \partial_\mu^b\partial_b\big[\bbvar{d}x^\nu]\partial^a_\nu=\bbvar{X}^b\bbwedge\partial_b\bbvar{X}^a.
\end{align}
In other words,\begin{equation}
\bbvar{d}\bbvar{X}^a = \bbvar{X}^b\bbwedge \nabla_b\bbvar{X}^a = \frac{1}{2}[\bbvar{X},\bbvar{X}]^a,\label{MCeq}
\end{equation}
where $\nabla_a$ denotes the covariant derivative for the metric $g_{ab}$.\medskip

Let us now proceed to write the extended pre-symplectic potential \eref{thetaext} in terms of the new variables, where the dressing is absorbed into a redefinition of the fields, as done in \eref{fieldredef} and \eref{Xdef} above. Taking into account the covariance \eref{covtheta} of the pre-symplectic potential and the commutators \eref{commut1} and \eref{commut2}, we immediately obtain 
\begin{equation}
\vartheta_{\mathit{ext}} = \vartheta -  \vartheta(\mathcal{L}_{\bbvar{X}})+\bbvar{X}\hook L=: \vartheta - \di q_\bbvar{X},\label{thetanew}
\end{equation}
where $q_{\bbvar{X}}$ denotes the charge aspect, which is a two-form on spacetime and one-form on field space, i.e.\ a section of $\bigwedge^2 T^\ast M\otimes T^\ast\mathcal{F}_{\mathit{ext}}$. It is then also useful to introduce the anticommuting\footnote{That is $Q_{\bbvar{X}}\bbwedge Q_{\bbvar{X}}=0$.}  Noether charge one-form on field space
\begin{equation}
Q_{\bbvar{X}} = \oint_{\partial\Sigma}q_{\bbvar{X}}\in T^\ast\mathcal{F}_{\mathit{kin}}.
\end{equation}

To proceed, we also need to introduce the pre-symplectic potential $\Theta_{\mathit{ext}}$ on the extended state space, whose exterior derivative defines the pre-symplectic two-form, i.e\
\begin{align}
\Theta_{\mathit{ext}} & = \int_\Sigma \vartheta_{\mathit{ext}},\\
\Omega_{\mathit{ext}}& = \bbvar{d}\Theta_{\mathit{ext}}.\label{Omex}
\end{align}
Note that for any two vector fields $\delta_1,\delta_2\in T^\ast \mathcal{F}_{\mathit{ext}}$, we thus have
\begin{equation}
\Omega_{\mathit{ext}}(\delta_1,\delta_2)=\delta_{1}\big[\Theta_{\mathit{ext}}(\delta_{2})\big]-\delta_{2}\big[\Theta_{\mathit{ext}}(\delta_{1})\big]-\Theta_{\mathit{ext}}\big([\delta_1,\delta_2]\big),
\end{equation}
where $[\cdot,\cdot]$ is the Lie bracket between vector fields on state space.

\subsection{Noether charges on the extended phase space}

\noindent The idea, which was developed in \cite{Ciambelli:2021vnn,PhysRevLett.128.171302, Freidel:2021dxw,Freidel:2021cjp,Speranza:2017gxd}, is to consider \emph{dressed diffeomorphisms} on the extended field space. A dressed diffeomorphism acts on metric and matter fields as well as on the dressing itself. Given a  vector field $\xi^a\in TM$, we consider thus the flow on field space,
\begin{equation}
g_{ab}\longmapsto\exp(\varepsilon\xi)^\ast g_{ab},\qquad \psi^I\longmapsto\exp(\varepsilon\xi)^\ast\psi^I.\label{diffeodef1}
\end{equation}
This flow is then compensated by a corresponding transformation of the dressing fields
\begin{equation}
\phi\longmapsto\exp(-\varepsilon\xi)\circ\phi=\phi\circ\exp(-\varepsilon\phi_\ast^{-1}\xi),\label{diffeodef2}
\end{equation}
such that the dressed fields $\phi^\ast g_{ab}$ and $\phi^\ast\psi^I$ are trivially invariant under \eref{diffeodef1} and \eref{diffeodef2}. 

Let us now consider what happens to this flow upon the field redefiniton
\begin{align}
    (\phi^\ast g_{ab}\longrightarrow g_{ab}, \phi^\ast\psi^I\longrightarrow \psi^I,\phi^{-1}_\ast\xi^a\longrightarrow\xi^a).\label{fieldredef2}
\end{align} Notice that this field redefinition has a natural and simultaneous action on the spacetime vector field $\xi^a\in TM$, sending $\xi^a$ into $\phi^{-1}_\ast\xi^a$. If we start out, in fact, with a \emph{field-independent} vector field $\xi^a$, i.e.\ $\bbvar{d}\xi^a=0$, the field redefinition \eref{fieldredef2}  maps $\xi^a$ into a field-dependent vector field on the extended state space. We shall see below how we are naturally led to consider such field dependent vector fields to render the charges integrable. After the field redefinition, the flow \eref{diffeodef1} and \eref{diffeodef2}  will only change the dressing fields, whereas its action on the metric and matter fields vanishes trivially. This flow lifts the vector field $\xi^a\in TM$ into a vector field $\delta_{\xi}^{\mathit{drssd}}$ on field space. Upon performing the field redefinition \eref{fieldredef2}, the components of this vector field are given by
\begin{align}
\delta^{\mathit{drssd}}_\xi[g_{ab}]=0,\quad\delta^{\mathit{drssd}}_\xi[\psi^I]=0,\quad\bbvar{X}^a(\delta^{\mathit{drssd}}_\xi)=-\xi^a.\label{deltadrssd}
\end{align}

Let us now identify the conditions necessary to make this vector field $\delta_\xi^{\mathit{drssd}}\in T\mathcal{F}_{\mathit{ext}}$ Hamiltonian. To this end, consider the interior product between the extended pre-symplectic two-form, which, upon performing the field redefinition \eref{fieldredef} and \eref{Xdef}, is given by \eref{thetanew} and \eref{Omex}, and the bivector $\delta\otimes\delta^{\mathit{drssd}}_\xi-\delta^{\mathit{drssd}}_\xi\otimes\delta$. A short calculation, see also \cite{Wald:1999wa}, gives
\begin{align}\nonumber
\Omega_{\mathit{ext}}(\delta,\delta^{\mathit{drssd}}_\xi)  = \delta\big[\Theta(\delta^{\mathit{drssd}}_\xi)\big]&-\delta^{\mathit{drssd}}_\xi\big[\Theta(\delta)\big]-\Omega\big([\delta,\delta^{\mathit{drssd}}_\xi]\big)+\\
&+\delta\big[Q_\xi\big]+\delta^{\mathit{drssd}}_\xi[Q_{\bbvar{X}(\delta)}]+Q_{\bbvar{X}([\delta,\delta^{\mathit{drssd}}_\xi])},\label{integralcharge}
\end{align}
where $\delta\in T\mathcal{F}_{\mathit{ext}}$ is a second and linearly independent vector field and
\begin{equation}
Q_\xi= \int_{\Sigma}\big(\vartheta(\mathfrak{L}_\xi)-\xi\hook L\big)=\oint_{\partial\Sigma}q_\xi\label{Noethercharge1}
\end{equation}
is the Noether charge.
 Given the definition \eref{deltadrssd} of the dressed diffeomorphisms $\delta^{\mathit{drssd}}_\xi$, the first line of equation \eref{integralcharge} vanishes trivially. The second line gives a non-trivial contribution
\begin{equation}
\delta^{\mathit{drssd}}_\xi[Q_{\bbvar{X}(\delta)}]=\bbvar{L}_{\delta_\xi^{\mathit{drssd}}}[Q_{\bbvar{X}(\delta)}]=Q_{\bbvar{L}_{\delta_\xi^{\mathit{drssd}}}[\bbvar{X}(\delta)]},
\end{equation}
where $\bbvar{L}_\delta[\cdot]=\delta\bbhook(\bbvar{d}\cdot)+\bbvar{d}(\bbhook\cdot)$ is the Lie derivative on the extended field space.
\begin{align}\nonumber
\bbvar{L}_{\delta_\xi^{\mathit{drssd}}}[\bbvar{X}^a(\delta)]&=(\bbvar{L}_{\delta_\xi^{\mathit{drssd}}}\bbvar{X}^a)(\delta)+\bbvar{X}^a\big([\delta_\xi^{\mathit{drssd}},\delta]\big)=\\\nonumber
&=(\bbvar{d}\bbvar{X}^a)(\delta_\xi^{\mathit{drssd}},\delta)-\delta[\xi^a]+\bbvar{X}^a\big([\delta_\xi^{\mathit{drssd}},\delta]\big)=\\
&=-[\xi,\bbvar{X}(\delta)]^a-\delta[\xi^a]+\bbvar{X}^a\big([\delta_\xi^{\mathit{drssd}},\delta]\big).
\end{align}
Thus
\begin{align}
\Omega_{\mathit{ext}}(\delta,\delta^{\mathit{drssd}}_\xi)  = \delta[Q_\xi]-Q_{\delta[\xi]-[\bbvar{X}(\delta),\xi]}.\label{intgrblty}
\end{align}
If the second term vanishes, the vector field $\delta_\xi^{\mathit{drssd}}$, defined in Equation \eref{deltadrssd} above, is Hamiltonian. The corresponding Hamiltonian is the Noether charge $Q_\xi$. The second term vanishes for any generic configuration on state space iff
\begin{equation}
\delta[\xi^a]=[\bbvar{X}(\delta),\xi]^a.\label{deltaxivec}
\end{equation}
This equation is satisfied for a specific class of field-dependent vector fields on spacetime, namely those that depend explicitly on the coordinates $\{x^\mu\}$ via their component functions $\xi^\mu(x)$,
\begin{equation}
\xi^a = \xi^\mu(x)\Big[\frac{\partial}{\partial x^\mu}\Big]^a\equiv \xi^\mu(x)\partial^a_\mu.
\end{equation}
In fact, such a vector field is field-dependent, because the four coordinate scalars $\{x^\mu\}$ have been added to the state space. To see that Equation \eref{deltaxivec} holds for such vector fields $\xi^a$ and field variations $\delta$, notice that
\begin{align}\nonumber
\delta[\xi^a] & = \delta[\xi^\mu(x)]\,\partial^a_\mu + 
\xi^\mu(x)\,\delta[\partial^a_\mu]=\\\nonumber
& = \delta[x^\nu]\,(\partial_\nu\xi^\mu)(x)\,\partial^a_\mu - 
\xi^\mu(x)\,\partial^b_\mu\,\delta[\di x^\nu_b]\,\partial^a_\nu=\\
& = \delta[x^\nu]\,(\partial_\nu\xi^\mu)(x)\,\partial^a_\mu - 
\xi^\mu(x)\,(\partial_\mu\delta[x^\nu])(x)\,\partial^a_\nu=[\delta[x],\xi]^a=[\bbvar{X}(\delta),\xi]^a,\label{Liexi}
\end{align}
where $[\cdot,\cdot]^a$ is the Lie bracket between vector fields on spacetime. Going back to \eref{intgrblty}, we thus see that the Noether charge \eref{Noethercharge1} generates the dressed diffeomorphism $\delta^{\mathit{drssd}}_\xi[\cdot]$ on the extended state space. From the definition \eref{deltadrssd} of the vector field, and the fact that these vector fields are Hamiltonian, we can now immediately infer the canonical commutation relations
\begin{align}
\big\{Q_\xi,Q_{\xi'}\big\}&=\delta_\xi^{\mathit{drssd}}[Q_{\xi'}]=Q_{\delta_\xi^{\mathit{drssd}}[{\xi'}]}=Q_{[\bbvar{X}(\delta_\xi^{\mathit{drssd}}),{\xi'}]}=-Q_{[\xi,{\xi'}]}.
\end{align}

\medskip{
Let us stop here and discuss the physcial significance of the approach outlined thus far. On the usual covariant phase space, the Komar charges for radial or time-like diffeomorphisms are not integrable. This is hardly surprising. Diffeomorphisms that move the boundary enlarge the system. They bring new data into the region that was in the exterior before. Since there is new data outside, there is no Hamiltonain for radial or time-like diffeomorphisms on the quasi-local phase space. Otherwise it would be possible to extend in a unique way initial data on a partial Cauchy surface into initial data on the entire Cauchy slice.

Upon performing a trivial field redefinition, we saw that the extended state space \cite{PhysRevLett.128.171302,Freidel:2021dxw} consists of the ordinary (undressed) fields in the bulk and the coordinate scalars $\{x^\mu:M\rightarrow \R^4\}$. The pre-symplectic structure on the extended state space is then carefully tuned in such a way that the conjugate momentum to the coordinates $\{x^\mu\}$ is the pull-back to $\Sigma$ of the exterior derivative of the Noether charge aspect, i.e.\  $p_\mu=\phi^\ast_\Sigma\di[ q_{\partial_\mu}]$. In this way, the pre-symplectic two-form is only changed by a boundary term at $\partial\Sigma$. Furthermore, all commutation relations between the new boundary fields and the dynamical fields in the interior vanish (upon the field redefinitions \eref{fieldredef} and \eref{Xdef}). Consider, for example, the total momentum charge with respect to the reference frame $\{x^\mu\}$, i.e.\ $P_\mu=\int_{\Sigma}p_\mu=Q_{\partial_\mu}$. 
Since the vector field $\delta^{\mathit{drssd}}_{\partial_\mu}[\cdot]=\{Q_{\partial_\mu},\cdot\}$ is Hamiltonian with respect to the extended pre-symplectic structure, and since $\delta^{\mathit{drssd}}_{\partial_\mu}$ annihilates all dynamical fields in the bulk, see \eref{deltadrssd}, the total momentum $P_\mu$ trivially commutes with all bulk degrees of freedom, i.e.\ 
 $\{P_\mu,g_{ab}\}=0$, $\{P_\mu,\psi^I\}=0$. The only-non vanishing Poisson bracket between $P_\mu$  and the elements of the extended state space is simply the Poisson bracket $\{P_\mu,x^\nu\}=\delta^\nu_\mu$. 
 
 While this is not a problem \emph{per se}, it does raise the question of how physically meaningful this extension of the phase space really is. There is no \emph{relational} change of matter and geometry relative to the hypersurface.  
 In the original construction due to \cite{PhysRevLett.128.171302}, dressed diffeomorphisms transform the fundamental fields, i.e.\ $g_{ab}\rightarrow\phi^\ast g_{ab}$, but they also deform the hypersurface, sending $\Sigma$ into $\phi^{-1}(\Sigma)$. From the perspective of an observer locked to $\Sigma$, the net effect is zero. Such dressed diffeomorphisms leave all covariant functionals of the metric at $\Sigma$ unchanged. Consider, for example, the total three-volume of $\Sigma$, i.e.\ the integral
\begin{equation}
\mathrm{Vol}[g_{ab},\Sigma] = \int_\Sigma \di^3 x\sqrt{\mathrm{det}(h_{ij})},\quad h_{ij} = g_{ab}\partial^a_i\partial^b_j,\label{threevol}
\end{equation}
where $\{x^i:i=1,2,3\}$ are coordinates intrinsic to $\Sigma$, $\partial^a_i\in T\Sigma$. Such a functional trivially Poisson commutes with all Noether charges $Q_\xi$ under the extended symplectic structure \cite{PhysRevLett.128.171302,Freidel:2021dxw}, i.e.\ $\{Q_\xi,\mathrm{Vol}\}=0$, even for those $\xi^a$ that are timelike. On the extended phase space, the Noether charge $Q_\xi$ does not behave like a physical time translation. A physical Hamiltonian should not preserve the total three-volume.
 Note that this constitutes an important difference between the dressed phase space approach and deparametrisation via physical reference frames. A material reference frame depends (in a complicated and non-linear manner) on the metric and matter fields and therefore does not  commute in general with the dynamical quantities in the bulk, see e.g.\ \eref{relational2}. 
 To put it simply, what is happening in \cite{PhysRevLett.128.171302,Freidel:2021dxw}  is that  the classical phase space is extended by adding new variables $x^\mu$ and $p_\mu$ and then carefully choosing a symplectic structure that allows us to identify $p_\mu$ with the Noether charge aspect, while, at the same time, all the newly added boundary variables (edge modes) trivially commute with all the dynamical fields in the bulk.
 }
\section{Gravitational subsystems and metriplectic geometry\label{sec3}}\noindent

In the following, we propose a different approach. We want to take seriously dissipation and treat the system as open.  Hence the Hamiltonian can not be conserved under its own flow. This can be formalised by replacing the symplectic structure by a metriplectic structure \cite{Morrison86,BKMR96,Fish2005}  with modified bracket, which captures dissipation (see \hyperref[app:metriplectic]{\darkg{Appendix B}} for a brief introduction to metriplectic geometry). The metriplectic structure consists of an extended symplectic two-form $\Omega_{\mathit{ext}}(\cdot,\cdot)\in T^\ast\mathcal{F}_{\mathit{ext}}\bigwedge T^\ast\mathcal{F}_{\mathit{ext}}$ and a symmetric bilinear form, namely a super-metric\footnote{Superspace is the space of fields.} $G(\cdot,\cdot)\in T^\ast\mathcal{F}_{\mathit{ext}}\bigotimes_{\mathit{sym}}T^\ast\mathcal{F}_{\mathit{ext}}$, which describes the interaction of the system with its environment. The resulting bilinear is then given by
\begin{equation}
K(\cdot,\cdot)=\Omega_{\mathit{ext}}(\cdot,\cdot)-G(\cdot,\cdot)\in T^\ast\mathcal{F}_{\mathit{ext}}\otimes T^\ast\mathcal{F}_{\mathit{ext}}.
\end{equation}

Given a functional $H:\mathcal{F}_{\mathit{ext}}\rightarrow\R$ on the extended state space, i.e.\ a functional $H[g_{ab},\psi^I,x^\mu]$ of the metric $g_{ab}$, the matter fields $\psi^I$ and the four coordinate scalars $x^\mu$, we say a vector field $\mathfrak{X}_H\in T\mathcal{F}_{\mathit{ext}}$ is Hamiltonian with respect to the metriplectic structure provided the following equation is satisfied,
\begin{equation}
\forall\delta\in T\mathcal{F}_{\mathit{ext}} : \delta[H]= K(\delta,\mathfrak{X}_H).\label{hamdissip}
\end{equation}
The new bracket between any two such functionals $H$ and $F$ on phase space is then given by
\begin{equation}
(H,F) = \mathfrak{X}_H[F].\label{brackdef}
\end{equation}
This bracket clearly satisfies the Leibniz rule in both arguments,
\begin{align}
(H_1H_2,F)&=H_1(H_2,F)+(H_1,F)H_2,\\
(H,F_1F_2)&=(H,F_1)F_2+F_1(H,F_2).
\end{align}

If there is dissipation, i.e.\ $G(\delta_1,\delta_1)\neq 0$,  the bracket will pick up a symmetric term such that the Hamiltonian will not be preserved under its own evolution, i.e.
\begin{equation}
(H,H) = - G(\mathfrak{X}_H,\mathfrak{X}_H).
\end{equation}
If, in addition, $H$ is the energy of the system, and the super-metric $G(\cdot,\cdot)$ is positive (negative) semi-definite, the system can only lose (gain) energy.\medskip

To apply metriplectic geometry to a gravitational subsystem in a finite domain $\Sigma$, we must identify the skew-symmetric  symplectic two-form $\Omega_{\mathit{ext}}$ and the super-metric $G(\cdot,\cdot)\in T^\ast\mathcal{F}_{\mathit{ext}}\bigotimes_{\mathit{sym}}$$T^\ast\mathcal{F}_{\mathit{ext}}$ that render the charges Hamiltonian. Our starting point is the familiar definition of the  Noether charge itself, i.e.\
\begin{equation}
Q_\xi = \int_\Sigma\Big(\vartheta(\mathfrak{L}_\xi)-\xi\hook L\Big),\label{Noethercharge0}
\end{equation}
where $\vartheta$ is the ordinary, undressed symplectic current and $L$ denotes the Lagrangian (a four-form on spacetime). In addition, $\mathfrak{L}_\xi\in T\mathcal{F}_{\mathit{ext}}$ is a vector field  on field space, whose components are given by the Lie derivative on the spacetime manifold, i.e.\
\begin{align}
\mathfrak{L}_\xi[g_{ab}] = \mathcal{L}_\xi g_{ab}=2\nabla_{(a}\xi_{b)},\qquad\mathfrak{L}_\xi[\psi^I] = \mathcal{L}_\xi\psi^I,\qquad\bbvar{X}^a(\mathfrak{L}_\xi) = \xi^a,\label{Lxidef}
\end{align}
where $\mathcal{L}_\xi$ is the Lie derivative of tensor fields on spacetime. Notice that this differs from the  dressed diffeomorphism $\delta_\xi^{\mathit{drssd}}$ which annihilates all fields in the bulk, see \eref{fieldredef} and \eref{deltadrssd}. 

On shell,\footnote{That is provided the field equations are satisfied.} the Noether charge \eref{Noethercharge0} is a surface integral,
\begin{equation}
Q_\xi=\int_{\Sigma}\di q_\xi=\oint_{\partial\Sigma}q_\xi.\label{Noethercharge2}
\end{equation}
We now want to identify the metriplectic structure that renders these charges the Hamiltonian generators of the field space vector field \eref{Lxidef}. Consider first the usual, undressed pre-symplectic two-form in the region $\Sigma$, i.e.\
\begin{equation}
\Omega = \int_\Sigma\bbvar{d}\vartheta \equiv \bbvar{d}\Theta.
\end{equation}
Given a vector field $\delta$ on field space, we then have
\begin{equation}
\Omega(\delta,\mathfrak{L}_\xi) = \delta\big[\Theta(\mathfrak{L}_\xi)\big]-\mathfrak{L}_\xi\big[\Theta(\delta)\big] -\Theta\big([\delta,\mathcal{L}_\xi]\big).\label{Omdeltaxi1}
\end{equation}
A standard calculation, see e.g.\ \cite{Wald:1999wa}, allows us to simplify the second term. First of all, we have
\begin{align}\nonumber
\mathfrak{L}_\xi\big[\Theta(\delta)\big]&=\int_\Sigma\mathfrak{L}_\xi\big(\vartheta[g_{ab},\psi^I;\delta g_{ab},\delta{\psi^I}](p)\big)=\\\nonumber
&=\int_\Sigma\int_M\Big((\mathfrak{L}_\xi g_{ab})(q)\frac{\delta\vartheta[g_{ab},\psi^I;h_{ab},\chi^I](p)}{\delta g_{ab}(q)}+(\mathfrak{L}_\xi h_{ab})(q)\frac{\delta\vartheta[g_{ab},\psi^I;h_{ab},\chi^I](p)}{\delta h_{ab}(q)}+\\
&\hspace{2em}+(\mathfrak{L}_\xi \psi^I)(q)\frac{\delta\vartheta[g_{ab},\psi^I;h_{ab},\chi^I](p)}{\delta \psi^I(q)}+(\mathfrak{L}_\xi \chi^I)(q)\frac{\delta\vartheta[g_{ab},\psi^I;h_{ab},\chi^I](p)}{\delta \chi^I(q)}\Big),\label{Lixithetadelta}
\end{align}
where $(\delta g_{ab},\delta\psi^I)\equiv(h_{ab},\chi^I)$ is a linearised solution of the field equations around $(g_{ab},\psi^I)$. The action of the vector field $\mathfrak{L}_\xi$ on the metric perturbation $h_{ab}$ yields
\begin{align}\nonumber
\mathfrak{L}_\xi h_{ab}&=[\mathfrak{L}_\xi,\delta]g_{ab}+\delta[\mathfrak{L}_\xi g_{ab}]
=[\mathfrak{L}_\xi,\delta]g_{ab}+\delta[\mathcal{L}_\xi g_{ab}]=\\\nonumber
&=[\mathfrak{L}_\xi,\delta]g_{ab}+[\delta,\mathcal{L}_\xi] g_{ab}+\mathcal{L}_\xi[\delta g_{ab}]=\\
&=[\mathfrak{L}_\xi,\delta]g_{ab}+\mathcal{L}_{\delta\xi} g_{ab}+\mathcal{L}_\xi[\delta g_{ab}].
\end{align}
In the same way, we also have
\begin{equation}
\mathfrak{L}_\xi \psi^I=[\mathfrak{L}_\xi,\delta]\psi^I+\mathcal{L}_{\delta\xi} \psi^I+\mathcal{L}_\xi[\delta \psi^I].
\end{equation}
Taking these results back to \eref{Lixithetadelta}, we obtain 
\begin{align}\nonumber
\mathfrak{L}_\xi[\Theta(\delta)]&=\Theta\big([\mathfrak{L}_\xi,\delta]\big)+\Theta(\mathfrak{L}_{\delta\xi})+\int_\Sigma\mathcal{L}_\xi[\vartheta(\delta)]=\\
&=\Theta\big([\mathfrak{L}_\xi,\delta]\big)+\Theta(\mathfrak{L}_{\delta\xi})+\oint_{\partial\Sigma}\xi\hook[\vartheta(\delta)]+\int_{\Sigma}\xi\hook\delta[L],\label{LiexiThetadelta}
\end{align}
where we used Stoke's theorem as well as the definition of the pre-symplectic potential in terms of the Lagrangian, i.e.\ the on-shell equation $\delta[L]=\di[\vartheta(\delta)]$.

Let us now return to \eref{Omdeltaxi1} above. Using the definition of the Noether charge \eref{Noethercharge1}, we obtain the well known result
\begin{equation}
\Omega(\delta,\mathfrak{L}_\xi)=\delta\big[Q_\xi\big]-Q_{\delta[\xi]}-\oint_{\partial\Sigma}\xi\hook\vartheta(\delta).\label{Omdeltaxi2}
\end{equation}

In the following, we shall restrict ourselves to a specific class of \emph{state dependent} vector fields on the extended state space.  The extended state space $\mathcal{F}_{\mathit{ext}}\ni(g_{ab},\psi^I,x^\mu)$ contains the coordinate functions  $x^\mu:M\rightarrow\R^4$. A vector field, given in terms of its $x^\mu$-coordinate representation, must be understood, therefore, as a state-dependent vector field,
\begin{align}
&\xi^a =\xi^\mu(x)\Big[\frac{\partial}{\partial x^\mu}\Big]^a\equiv \xi^\mu(x)\partial^a_\mu,\\
&\delta[\xi^\mu]=\int_M\delta[x^\nu]\frac{\delta}{\delta x^\nu}\xi^\mu\big(x(p)\big)=\delta[x^\nu]\big(\partial_\nu\xi^\mu\big)\big(x(p)\big).\label{statedepxi}
\end{align}
Note that $\xi^a$ depends as a functional on $x^\mu: M \to \mathbb{R}^4$, but there is no functional dependence on $g_{ab}$ or $\psi^I$. This way, the functional differential $\bbvar{d}\xi^a$ of any such vector field returns the Lie derivative on spacetime with respect to the Maurer--Cartan form $\bbvar{X}^a$, i.e.\ $\xi^a = \bbvar{X}^a(\mathfrak{L}_\xi)$ and $\delta[\xi^a] = [\bbvar{X}(\delta),\xi]^a$ (cf.\ Equation \eref{Liexi} above). For any such specific \emph{state-dependent} vector field, we can rewrite Equation \eref{Omdeltaxi2} as
\begin{align}
    \delta[Q_\xi] = \Omega(\delta,\mathfrak{L}_\xi) + Q_{[\bbvar{X}(\delta),\bbvar{X}(\mathfrak{L}_\xi)]} + \oint_{\partial \Sigma}\bbvar{X}(\mathfrak{L}_\xi)\hook \vartheta(\delta).
\end{align}
Comparing this equation with the definition of Hamiltonian vector fields for a dissipative system, i.e.\ Equation \eref{hamdissip}, and demanding that the Lie derivative $\mathfrak{L}_\xi\in T\mathcal{F}_{\mathit{ext}}$ be the Hamiltonian vector field of the Noether charge $Q_\xi$, we are led to the following definition: a vector field $\mathfrak{X}_H\in T\mathcal{F}_{\mathit{kin}}$ is Hamiltonian, if there exists a functional $H:\mathcal{F}_{\mathit{ext}}\rightarrow\R$ on state space, such that for all vector fields $\delta\in T\mathcal{F}_{\mathit{ext}}$ the following condition is satisfied,
\begin{equation}
\delta[H] = \Omega(\delta,\mathfrak{X}_H)+Q_{[\bbvar{X}(\delta),\bbvar{X}(\mathfrak{X}_H)]}+\oint_{\partial\Sigma}\bbvar{X}(\mathfrak{X}_H)\hook\vartheta(\delta)\equiv K(\delta,\mathfrak{X}_H).\label{deltaH}
\end{equation}
The new bracket between any two such functionals is then given by Equation \eref{brackdef}. Moreover, we are now ready to identify the metrisymplectic structure that renders the charges integrable, i.e.\
\begin{equation}
K(\cdot,\cdot) = \Omega_{\mathit{ext}}(\cdot,\cdot)-G(\cdot,\cdot).\label{Leibniz1}
\end{equation}
The skew-symmetric part  defines the extended symplectic two-form
\begin{equation}
\Omega_{\mathit{ext}}(\delta_1,\delta_2)=-\Omega_{\mathit{ext}}(\delta_2,\delta_1)=\Omega(\delta_1,\delta_2)+Q_{[\bbvar{X}(\delta_1),\bbvar{X}(\delta_{2})]}+\oint_{\partial\Sigma}\bbvar{X}(\delta_{[1})\hook\vartheta(\delta_{2]}),\label{Leibniz2}
\end{equation}
where square brackets around the indices stand for anti-symmetrisation, i.e.\ $(\alpha\bbwedge\beta)(\delta_1,\delta_2)=2\alpha(\delta_{[1})\beta(\delta_{2]})=\alpha(\delta_1)\beta(\delta_2)-(1\leftrightarrow 2)$ for all $\alpha,\beta\in T^\ast\mathcal{F}_{\mathit{kin}}$. The symmetric part, on the other hand, determines the super-metric
\begin{equation}
G(\delta_1,\delta_2)=-\oint_{\partial\Sigma}\bbvar{X}(\delta_{(1})\hook\vartheta(\delta_{2)}),\label{Leibniz3}
\end{equation}
where the round brackets around the indices stand for symmetrisation, i.e.\ $(\alpha\otimes\beta)(\delta_{(1},\delta_{2)})=\frac{1}{2}\big(\alpha(\delta_1)\beta(\delta_2)+\alpha(\delta_2)\beta(\delta_1)\big)$. Note that the super-metric $G(\cdot,\cdot)$ is a boundary term. This is consistent with our physical intuition that the interaction of an open system with its environment takes place at the boundary.\medskip

Let us briefly summarise.  We introduced a new bracket $(\cdot,\cdot)$ on state space that turns the covariant phase space into a metriplectic manifold. This bracket is a generalisation of the Poisson bracket. It takes into account dissipation and renders the vector field $\mathfrak{L}_\xi[\cdot]$, defined in \eref{Lxidef}, integrable. The corresponding Hamiltonian is the Noether charge,
\begin{equation}
(Q_\xi,g_{ab})=\mathcal{L}_\xi g_{ab},\quad (Q_\xi,\psi^I)=\mathcal{L}_\xi\psi^I,\quad(Q_\xi,x^\mu)=\xi^\mu.
\end{equation}
These results are only possible at the expense of changing the bracket. Neither does the new bracket satisfy the Jacobi identity nor is it skew-symmetric. The symmetric part describes dissipation. The skew-symmetric part defines the usual Poisson bracket on the extended phase space. \medskip

Let us add a few further observations. We built the Leibniz bracket in such a way that the Noether charge generates the Hamiltonian vector field \eref{Lxidef}. Given two state dependent vector fields $\xi_1^a=\xi_1^\mu(x)\partial^a_\mu$ and $\xi^a_2=\xi_2^\mu(x)\partial^a_\mu$ that satisfy Equation \eref{statedepxi}, we can now also obtain immediately the new bracket between two such charges, i.e.\
\begin{align}
(Q_{\xi_1},Q_{\xi_2}) & = \mathfrak{L}_{\xi_1}[ Q_{\xi_2}]=\oint_{\partial\Sigma}{\xi_1}\hook(\di q_{{\xi_2}})=\oint_{\partial\Sigma}\Big({\xi_1}\hook\vartheta(\mathfrak{L}_{\xi_2})-{\xi_1}\hook{\xi_2}\hook L\Big).
\end{align}
In the same way, we obtain the Leibniz bracket of the Noether charge with itself,
\begin{align}
(Q_\xi,Q_\xi) & = -G(\mathfrak{L}_\xi,\mathfrak{L}_\xi)=\oint_{\partial{\Sigma}}\xi\hook\vartheta(\mathfrak{L}_\xi).
\end{align}
If the vector field $\xi^a\in TM$ lies tangential to the corner, i.e.\ $\xi^a\in T(\partial\Sigma)$, the charge is conserved under its own Hamiltonian flow. Intuitively, this must be so, because the resulting diffeomorphism maps the corner \emph{relative} to the metric into itself. Hence, there is no relational change.  On the other hand, a generic diffeomorphism that moves the boundary \emph{relative} to the metric, will not preserve its own Hamiltonian if there is flux, i.e.\ $\xi\hook\vartheta(\mathfrak{L}_\xi)\big|_{\partial\Sigma}\neq0$.

\section{Outlook and Discussion}

\noindent In this work, we discussed two different approaches towards describing the phase space of a gravitational subsystem localised in a compact region of space: the extended covariant phase space approach due to \cite{PhysRevLett.128.171302,Freidel:2021dxw} as well as a new geometrical framework based on metriplectic geometry \cite{Morrison86,BKMR96,Fish2005}. The former is focused on obtaining integrable charges for diffeomorphisms, including \emph{large diffeomorphisms} that change the boundary.  To achieve this, the phase space is extended. Embedding fields $x^\mu:M\rightarrow \R^4$ are added to phase space and the pre-symplectic structure is modified accordingly. The key result \cite{Ciambelli:2021vnn,PhysRevLett.128.171302,Freidel:2021dxw,Freidel:2021cjp,Speranza:2017gxd} is algebraic: On the extended phase space, the Komar charges close under the Poisson bracket. This yields a new Hamiltonian representation of the Lie algebra of vector fields on spacetime. However, this comes at the cost of weakening the physical interpretation of the charges. Upon performing a trivial field redefinition, we saw that the charges commute with all bulk degrees of freedom. The Hamiltonian vector field of the charges only shifts the embedding coordinates at the boundary. Put differently, on the extended phase space \cite{PhysRevLett.128.171302,Freidel:2021dxw}, the Komar charge generates diffeomorphisms of the metric and the matter fields, but such change is always made undone by a deformation of the hypersurface $\Sigma$. For an observer locked to $\Sigma$, the net effect is zero. 
 
The metriplectic approach provides a new perspective on how to obtain meaningful charges on phase space. Once again, the Komar charges are rendered Hamiltonian, yet the bracket is different. Instead of the Poisson bracket, we now have a Leibniz bracket $(\cdot,\cdot)$. The resulting Hamiltonian vector field $(Q_\xi,\cdot)$ generates the full non-linear dynamics in the interior of $\Sigma$ while accounting for the interaction of the system with its environment. This is achieved by replacing the usual pre-symplectic structure on phase space with the metriplectic structure commonly used in the context of dissipative systems \cite{Morrison86,BKMR96,Fish2005}. The main difference to the extended phase space approach is that the Leibniz bracket will no longer provide a representation of the diffeomorphism group, i.e.\ there is an anomaly $(Q_{\xi_1},Q_{\xi_2})\neq -Q_{[\xi_1,\xi_2]}$. The extra terms account for dissipation and flux.

What both approaches have in common is that they give a rigorous meaning to the Komar charges on phase space. Therefore, they both face the same problem of what is the physical interpretation of these charges. To compute the Komar charge on state space, we need three inputs: a choice of hypersurface $\Sigma$, a vector field $\xi^a \in TM$, and a solution to the field equations. This leaves a lot of functional freedom. At finite distance, it is difficult to explain how such charges are connected to physical observables such as energy, momentum, angular momentum. Given the metric and the Cauchy hypersurface, one is left with infinitely many choices for the vector field $\xi^a$. 
It is unclear which $\xi^a$ gives rise to energy, which to momentum, and which to angular momentum.  However, this is just a reflection of background invariance. If the theory is background invariant, there is an infinite-dimensional group of gauge symmetries (diffeomorphisms). These infinitely many gauge symmetries, give rise to infinitely many charges, hence the vast functional freedom in defining the quasi-local charges. A second potential criticism is that the first derivative of the Komar charge, accounting for flux, does not vanish in Minkowski space. This may seem counterintuitive at first. Minkowski space is empty and thus no flux expected. However, for a given choice of vector field $\xi^a$, the flux of the Komar charge  depends not only on radiative data, but also on kinematical data. The kinematical data is the choice of boundary $\partial\Sigma$ and the choice of vector field $\xi^a$. To probe the presence of curvature and distinguish purely kinematical flux from physical flux due to gravitational radiation, it seems necessary to add further derivatives, e.g.\ $(Q_\xi,(Q_\eta,Q_\tau))$. Future research will be necessary to clarify the physical significance of such nested brackets and their algebraic properties in terms of e.g.\ the Jacobiator $J(\xi,\eta,\tau) = (Q_\xi,(Q_\eta,Q_\tau))+(Q_\eta,(Q_\tau,Q_\xi))+(Q_\tau,(Q_\xi,Q_\eta))$. 

Another important avenue for future research concerns black holes. Black holes have an entropy and there is a notion of energy and temperature. The outside region, connected to asymptotic infinity, defines a dissipative  system: Radiation can fall into the black hole, but nothing comes out.  The metriplectic approach is tailor-made to study such thermodynamical systems out of equilibrium, to investigate chaos, stability, and dissipation. Entropy production and energy loss are captured by the super-metric $G(\cdot,\cdot)$ on metriplectic space.

Finally, let us briefly comment on the implications for quantum gravity. In metriplectic geometry, the Liouville theorem is violated. The volume two-form on phase space is no longer conserved under the Hamiltonian flow $(Q_\xi,\cdot)$. An analogous statement should be possible at the quantum level. Evolution should be now governed by a non-unitary dynamics, e.g.\ a flow-equation consisting of an anti-symmetric commutator, representing the unitary part, and a symmetric Lindbladian describing the radiation.\vspace{-0.4em}

\section*{Acknowledgements}
\noindent We acknowledge financial support by the Austrian Science Fund (FWF) through BeyondC (F7103-N48), the Austrian Academy of Sciences (\"OAW) through the project ``Quantum Reference Frames for Quantum Fields'' (ref.~IF 2019\ 59\ QRFQF), and of the ID 61466 grant from
the John Templeton Foundation, as part of the ``Quantum Information Structure of Spacetime (QISS)'' project (qiss.fr).\vspace{-0.4em} 

\providecommand{\href}[2]{#2}\begingroup\raggedright\endgroup

\appendix
\section[Notation and conventions]{Notation and conventions}
\paragraph{- Index Notation} We use a hybrid notation. $p$-form indices are often suppressed, but tensor indices are kept. Indices $a,b,c,\dots$ from the first half of the alphabet are abstract indices on tangent space. Indices $\mu,\nu,\rho,\dots$ from the second half of the Greek alphabet refer to coordinate charts $\{x^\mu:U_\mu\subset M\rightarrow\R^4\}$. 
\paragraph{- Spacetime} We are considering a spacetime manifold $M$, with signature $(-$$+$$+$$+)$, metric $g_{ab}$ and matter fields $\psi^I$ that satisfy the Einstein equations $R_{ab}-\tfrac{1}{2}g_{ab}R=8\pi G\,T_{ab}$ and the field equations for the matter content. On this manifold, we have several natural derivatives. $\nabla_a$ denotes the usual (metric compatible, torsionless) derivative, $\mathcal{L}_\xi$ is the Lie derivative for a vector field $\xi^a \in TM$, and \qq{$\di$} denotes the exterior derivative, i.e.\ $(\di\omega)_{a_1\dots a_{p+1}}=(p+1)\nabla_{[a_1}\omega_{a_2\dots a_{p+1}]}$. If $\omega$ is a $p$-form on $M$, the Lie derivative satisfies $\mathcal{L}_\xi\omega=\di(\xi\hook \omega)+\xi\hook(\di\omega)$, where $(\xi\hook\omega)(\eta,\dots)=\omega(\xi,\eta,\dots)=\omega_{ab\dots}\xi^a\eta^b\cdots$ is the interior product. If applied to a vector field, the Lie derivative acts via the Lie bracket $\mathcal{L}_\xi\eta^a =[\xi,\eta]^a=\xi^b\nabla_b\eta^a-\eta^b\nabla_b\xi^a$.
\paragraph{- Field space} Field space $\mathcal{F}$ is the state space of the solutions of the field equations. For simplicity, we always go \emph{on-shell;} otherwise, we would need to constantly carry around terms that are constrained to vanish.  
As for the differential calculus on $\mathcal{F}$, the following notation is used. Linearised solutions $\delta[g_{ab}]=:h_{ab}$, $\delta[\psi^I]=:\chi^I$ define tangent vectors $\delta\in T\mathcal{F}$ on field space. If, in fact, $(g^{(\varepsilon)}_{ab},\psi^I_{(\varepsilon)})$ is a smooth one-parameter family of solutions to the field equations, through the point on field space $(g_{ab},\psi^I)=(g^{(\varepsilon)}_{ab},\psi^I_{(\varepsilon)})\big|_{\varepsilon=0}$, we set
\begin{align}
\delta[g_{ab}]&=\frac{\di}{\di\varepsilon}\Big|_{\varepsilon=0}g^{(\varepsilon)}_{ab},\label{gvar}\\
\delta[\psi^I]&=\frac{\di}{\di\varepsilon}\Big|_{\varepsilon=0}\psi^I_{(\varepsilon)}.
\end{align}
To distinguish the differential calculus on field space from the differential calculus on spacetime, we use a double stroke notation wherever necessary: $\bbvar{d}$ is the exterior derivative on field space, $\bbwedge$ denotes the wedge product between differential forms on $\mathcal{F}$, and $\bbhook$ is the interior product. If $F:\mathcal{F}\rightarrow \R$ is a differentiable functional on state space, we may thus write,
\begin{equation}
\delta[F]=\delta\bbhook\bbvar{d}F.
\end{equation}
If, in addition, $\delta$ is a vector field on field space, and $\Xi$ is a $p$-form on field space, the Lie derivative on state space will satisfy the familiar identities

\begin{align}
\bbvar{L}_\delta[\Xi] & = \delta\bbhook\big(\bbvar{d}[\Xi]\big)+\bbvar{d}\big(\delta\bbhook[\Xi]\big),\\
\bbvar{L}_\delta[\bbvar{d}\Xi] & = \bbvar{d}[\bbvar{L}_\delta\Xi].
\end{align}

\paragraph{- Komar charge} For the Einstein--Hilbert action with matter action $L_{\mathit{matter}}[g_{ab},\psi^I,\nabla_a\psi^I]$, the pre-symplectic potential is given by
\begin{equation}
\Theta(\delta)= \int_\Sigma\vartheta(\mathfrak{L}_\xi)=\frac{1}{16\pi G}\int_\Sigma d^3v_a\big(\nabla_b h^{ab}-\nabla^a\ou{h}{b}{b}\big)+\int_\Sigma d^3v_a\frac{\partial L_{\mathit{matter}}}{\partial(\nabla_a\psi^I)}\chi^I,
\end{equation}
where $(h_{ab},\chi^I)\equiv(\delta g_{ab},\delta\psi^I)$ solves the linearised field equations and $d^3v_a$ is the directed volume element (a tensor-valued $p$-form). More generally,
\begin{equation}
d^{p}v_{a_1\dots a_{4-p}}=\frac{1}{p!}\varepsilon_{a_1\dots a_{4-p}b_{1}\dots b_p}\partial^{b_1}_{\mu_1}\cdots\partial^{b_p}_{\mu_p}\di x^{\mu_1}\wedge\dots\wedge\di x^{\mu_p}.
\end{equation}
On shell, the Noether charge is given by the Komar formula
\begin{equation}
Q_\xi =\int_\Sigma\big(\vartheta(\mathfrak{L}_\xi)-\xi\hook L\big)=-\frac{1}{16\pi G}\oint_{\partial \Sigma}d^2v^{ab}\,\nabla_{[a}\xi_{b]},
\end{equation}
where $\mathfrak{L}_\xi$ is the Lie derivative and $L=d^4v\,\big(({16\pi G})^{-1}R[g,\partial g,\partial^2 g]+L_{\mathit{matter}}[g,\psi,\nabla\psi]\big)$ is the total Lagrangian.

\section[Metriplectic space and dissipation]{Metriplectic space and dissipation}\label{app:metriplectic}
\noindent  In this section, we briefly review the formalism of metriplectic systems \cite{Morrison86,BKMR96,Fish2005} as an extension of the framework for Hamiltonian systems. To simplify the exposition, we restrict ourselves in this appendix to a finite-dimensional system.  The generalisation to field theory is straightforward.

For \emph{Hamiltonian systems}, the phase space is characterised by the equations of motion
\begin{align}
    \frac{\di}{\di t}f = \{H,f\}
\end{align}
defined in terms of the anti-symmetric Poisson bracket
\begin{align}
    \{f,g\} = \omega^{ij}\frac{\partial f}{\partial z^i}\frac{\partial g}{\partial z^j},
\end{align}
where $\omega^{ij}=-\omega^{ji}$ is the inverse of the symplectic two-form
\begin{equation}
\omega=\frac{1}{2}\omega_{ij}\,\di z^i\wedge\di z^j,\quad \di\omega=0,\quad\omega^{jm}\omega_{im}=\delta^j_i,
\end{equation}
 and $z^i$, $i=1\dots 2N$ are coordinates on phase space. Analogously, one can define a \emph{metric system} through the equations of motion
\begin{align}
    \frac{\di}{\di t}f = \llbrack S, f \rrbrack
\end{align}
and a \emph{symmetric} bracket
\begin{align}
     \{\hspace{-0.23em}| f,g|\hspace{-0.23em}\} = g^{ij}\frac{\partial f}{\partial z^i}\frac{\partial g}{\partial z^j},
\end{align}
with inverse metric tensor $g^{ij} = g^{ji}$ and line element
\begin{equation}
d s^2 =g_{ij}\,\di z^i\otimes \di z^j.
\end{equation}
If one requires, in addition, that $g^{ij}$ is positive-definite it follows that
\begin{align}
    \frac{\di}{\di t}S = g^{ij}\frac{\partial S}{\partial z^i}\frac{\partial S}{\partial z^j} \geq 0,
\end{align}
i.e.\ $S$ only increases over time and finds its interpretation as a form of entropy. Finally, one obtains a \textit{metriplectic system} by combining the two brackets to form the Leibniz bracket
\begin{align}
     (f,g) = \{f,g\} \pm \llbrack f,g\rrbrack
\end{align}
where the relative sign depends on the conventions and on which thermodynamical potential generates the evolution (e.g.\ internal energy, free energy, entropy).

Note that this bracket lives up to its name and satisfies the Leibniz rule in either argument
\begin{align}
(f,gh) & = (f,g)h+g(f,h),\\
(fg,h) & = f(g,h)+(f,h)g,
\end{align}
for all phase-space functions $f,g,h$. Based on this bracket, there are different ways to define the equations of motion. On the one hand, one can introduce a generalised free energy $F = H - TS$ to generate the flow of the metriplectic system, in which case the Hamiltonian is conserved and the dissipation captured by an increase in entropy \cite{Morrison86}. Alternatively one can use the Hamiltonian itself to generate the time evolution through
\begin{align}
    \frac{\di}{\di t}f = (H,f).
\end{align}
In this case, the Hamiltonian is no longer conserved since $(H,H) \neq 0$, due to the symmetric part of the Leibniz bracket, and captures directly the loss or gain of energy through dissipation, depending on the sign of $\llbrack H,H\rrbrack$, \cite{BKMR96,Fish2005}. 

Furthermore, just as we can associate Hamiltonian vector fields ${X}_f$ to each function $f$ on phase space given a symplectic two-form $\omega$ through $\delta[f] = \omega(\delta,{X}_f)$ for all variations $\delta$, we can define Hamiltonian vector fields ${X}_f$ with respect to the metriplectic structure as
\begin{align}
   \forall \delta : \delta f = \omega(\delta,{X}_f) \pm g(\delta,{X}_f),
\end{align}
provided the bilinear $k(\cdot,\cdot)=\omega(\cdot,\cdot)\pm g(\cdot,\cdot)$ is non-degenerate.\footnote{In gauge theories, $k(\cdot,\cdot)$ will have non-trivial null vectors. A gauge fixing amounts to taking the pull-back to a submanfiold, where $k(\cdot,\cdot)$ is non-degenerate.} Given the metric and symplectic two-form, we define the Leibniz bracket 
\begin{align}
    (f,g) = \omega({X}_f,{X}_g) \pm g({X}_f,{X}_g) = {X}_f[g].
\end{align}
Stricly speaking, there are two Leibniz vector fields, namely a right Leibniz vector field ${X}_H^r[f] = (f,H)$ and a left Leibniz vector field $X_H^\ell[f] = - (H,f)$. These definitions are the same for antisymmetric brackets but no longer so for the Leibniz bracket.

\end{document}